\documentclass[journal]{IEEEtran}
\usepackage{amsmath,amsfonts}
\usepackage{siunitx}
\usepackage{algorithmic}
\usepackage{array}
\usepackage{multirow}
\usepackage[caption=false]{subfig}
\usepackage{textcomp}
\usepackage{stfloats}
\usepackage{url}
\usepackage{verbatim}
\usepackage{graphicx}
\usepackage{xcolor}
\usepackage[noadjust]{cite}
\usepackage[inline]{enumitem}
\usepackage{xspace}
\usepackage{booktabs}

\usepackage{footnote}
\makesavenoteenv{tabular}
\makesavenoteenv{table}

\graphicspath{{figs/}}

\newcommand{\hsi}{H-Si(100)-2$\times$1\@\xspace}
\newcommand{\figref}[1]{Fig.~\ref{#1}}
\newcommand{\tabref}[1]{Table~\ref{#1}}
\newcommand{\secref}[1]{Section~\ref{#1}}
\newcommand{\eref}[1]{Eq.~\eqref{#1}}

\newcommand{\aka}{a.k.a.\@\xspace}
\newcommand{\etal}{\textit{et~al.}}

\DeclareSIUnit\angstrom{\text {\AA}}

\def\BibTeX{{\rm B\kern-.05em{\sc i\kern-.025em b}\kern-.08em
    T\kern-.1667em\lower.7ex\hbox{E}\kern-.125emX}}
\usepackage{balance}

\begin{document}
\title{Simulating Charged Defects in Silicon Dangling Bond Logic Systems to Evaluate Logic Robustness}

\author{Samuel~S.~H.~Ng,    %
  Jeremiah Croshaw,         %
  Marcel Walter,            %
  Robert Wille,             %
  Robert Wolkow,            %
  Konrad~Walus              %
  \thanks{This work was supported by the Natural Sciences and Engineering Research Council of Canada under Grant \mbox{RGPIN-2022-04830}.}%
  \thanks{S.~S.~H.~Ng and K.~Walus were with the Department of Electrical and Computer Engineering, University of British Columbia, Vancouver, BC, Canada (\mbox{e-mail}: \mbox{samueln@ece.ubc.ca}; \mbox{konradw@ece.ubc.ca}); J.~Croshaw and R.~Wolkow were with the Department of Physics, University of Alberta, Edmonton, AB, Canada and Quantum Silicon Inc., Edmonton, AB, Canada (\mbox{e-mail}: \mbox{croshaw@ualberta.ca}; \mbox{rwolkow@ualberta.ca}); M.~Walter and R.~Wille were with the Chair for Design Automation, Technical University of Munich, BY, Germany (e-mail: \mbox{marcel.walter@tum.de}, \mbox{robert.wille@tum.de}). R.~Wille was also with the Software Competence Center Hagenberg (SCCH) GmbH, Hagenberg, O\"O, Austria.}%
}

\maketitle

\begin{abstract}
Recent research interest in emerging logic systems based on quantum dots has been sparked by the experimental demonstration of nanometer-scale logic devices composed of atomically sized quantum dots made of silicon dangling bonds (SiDBs), along with the availability of SiQAD, a computer-aided design tool designed for this technology.
Latest design automation frameworks have enabled the synthesis of SiDB circuits that reach the size of $\mathbf{32\times10^3}\,\text{\textbf{nm}}^\mathbf{2}$---orders of magnitude more complex than their hand-designed counterparts.
However, current SiDB simulation engines do not take defects into account, which is important to consider for these sizable systems. This work proposes a formulation for incorporating fixed-charge simulation into established ground state models to cover an important class of defects that has a non-negligible effect on nearby SiDBs at the $\mathbf{10}\,\text{\textbf{nm}}$ scale and beyond. The formulation is validated by implementing it into SiQAD's simulation engine and computationally reproducing experiments on multiple defect types, revealing a high level of accuracy.
The new capability is applied towards studying the tolerance of several established logic gates against the introduction of a single nearby defect to establish the corresponding minimum required clearance. These findings are compared against existing metrics to form a foundation for logic robustness studies.
\end{abstract}


\section{Introduction}

\IEEEPARstart{A}{lternative} logic implementations have been of immense interest as the further shrinkage of transistors becomes increasingly challenging and costly. Field-coupled nanocomputing is one such alternative---it defines a class of devices that operate based on field interactions with promises of ultra-low power and high frequency operation at the atomic scale \cite{lent1997device, lent2003molecular, bernstein2005magnetic, walus2004qcadesigner}.
One such implementation comes in the form of quantum dots made of silicon dangling bonds (SiDBs) on the hydrogen passivated silicon (100) 2$\times$1 surface (\hsi), with the experimentally demonstrated capability to implement an OR gate that spans the length of $<\SI{10}{\nm}$ \cite{huff2018binary}.
This groundbreaking experimental demonstration, combined with latest developments in computer-aided design (CAD) capabilities offered by SiQAD and compatible simulators \cite{ng2020siqad, chiu2020poissolver, drewniok2023quicksim}, has spurred wide ranging research interests in the technology including the proposal of gate designs \cite{ng2020siqad, chiu2020thes, bahar2020atomic, vieira2021novel, vieira2022threeinput}, design automation support from the fiction framework \cite{walter2019fiction,walter2022hexagons}, an automated quantum dot gate design tool based on reinforcement learning \cite{lupoiu2022automated}, and evaluations of future applications \cite{ng2020thes, chiu2020thes, ng2023ablueprint}.
However, current simulation capabilities are designed for an ideal physical environment consisting of a perfect, i.e., defect-free, silicon monocrystal substrate. While this can be justified at the individual logic gate level as gates are small enough to fit within a relatively defect-free region for experimentation purposes, the effect of defects cannot be neglected especially in light of recent design automation works that have synthesized SiDB circuits reaching the size of $\SI{32e3}{\nm^2}$ \cite{walter2022hexagons}.

Multiple types of defects may be found on the \hsi surface, including electrically charged and neutral specimens \cite{croshaw2020atomic}.
This work proposes a formulation to capture the effects of fixed-charge defects in simulation. We implement the formulation in SimAnneal, an efficient three-state ground state charge configuration simulator offered by SiQAD, and validate the simulation results with previous experimental work \cite{huff2019electrostatic}.
We further demonstrate the relevance of the added capability by using it to study the robustness of SiDB logic gates against the introduction of charged defects and compare the results with the gates' tolerance against changes in physical parameters, \aka the operational domain \cite{ng2020siqad}.

An overview of the physical backgrounds and logic design efforts on SiDB logic is provided in \secref{sec:background};
\secref{sec:methodology} describes the formulation of fixed-charge defects in the context of established ground state models;
\secref{sec:verification} verifies the defect simulation capabilities by comparing results against physical experiments \cite{huff2019electrostatic};
\secref{sec:application} demonstrates the application of the methodology in the analysis of the resilience of logic gates against charged defects nearby;
and lastly, \secref{sec:conclusion} concludes our efforts and discusses future work.

\section{Background}\label{sec:background}

In this section, we provide a key summary of existing literature on the fabrication of SiDBs and utilization in logic representation, the characterization of atomic defects that exert localized effects, and the evaluation of the stability of logic networks in non-ideal systems.

\subsection{Fabrication and Logic Representation}

\begin{figure}
  \centering
  \includegraphics[width=.5\linewidth]{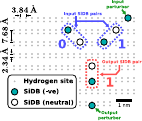}
  \caption{Illustration of the \hsi surface with hydrogen sites in gray, neutral SiDBs as hollow circles, and negatively charged SiDBs as teal-filled circles. The separation distances of hydrogen sites are given in the top left of the figure. The right side of the figure illustrates a reproduction of an OR gate from \cite{ng2020siqad} in logic $01$ configuration. The direction of logic flows from top to bottom. The input SiDB-pairs take on logic $0$ by default and is set to logic $1$ in the presence of input perturbers. The existence of an output perturber sets the default output SiDB-pair logic state to $0$, with the logic state being pushed to $1$ when one or both inputs are set to $1$.}
  \label{fig:background}
\end{figure}

SiDBs can be fabricated on the \hsi surface by the removal of hydrogen atoms using a scanning probe \cite{achal2018lithography}; they can also be erased by repassivating the SiDB \cite{huff2017atomic}. The periodicity of the lattice structure guarantees atomically accurate hydrogen locations on the surface as illustrated in \figref{fig:background}.
SiDBs were found to behave like quantum dots as their energetic charge transition levels fall within the bulk band gap, allowing them to host 0, 1, or 2 electrons corresponding to positive, neutral, and negative charge states \cite{pitters2011charge, taucer2014singleelectron}.
In \cite{huff2018binary}, Huff \etal{}~experimentally demonstrated logic cells, wires, and an OR gate that represents logic states via the location of charges shared between pairs of SiDBs as shown in \figref{fig:background}. This representation is later denoted binary-dot logic (BDL).

Further exploration into SiDB logic implementations came in the form of computer-aided studies with the use of SiQAD \cite{ng2020siqad}.
Y-shaped BDL gates implementing various common logic functions were proposed \cite{ng2020siqad, walter2022hexagons}, as were T/$+$-shaped BDL gates \cite{bahar2020atomic, vieira2021novel, vieira2022threeinput} and gates that employ alternative logic representations \cite{ng2020thes}, exemplifying the flexibility of the SiDB logic platform. Gate libraries and design automation frameworks have also been developed \cite{walter2022hexagons}. However, the effect of defects on these logic gates have yet to be extensively investigated.

\subsection{Experimental Characterization of Defects}

\begin{figure}
  \centering
  \includegraphics[width=.6\linewidth]{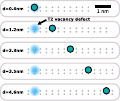}
  \caption{Illustration of the T2 silicon vacancy defect characterization experimental setup from \cite{huff2019electrostatic}. The defect is at a depth of \SI{0.8}{\angstrom} below the surface and is represented by a blue halo. At each experimental step, a probe-SiDB is created at some distance from the defect, voltage sweeping experiments are performed on the SiDB, then the SiDB is removed and recreated at a different distance. The Coulombic effects exerted by the defect on the probe-SiDB weakens as distance increases, allowing parameters to be fitted.}
  \label{fig:exp-setup}
\end{figure}

Various types of material imperfections at the atomic scale have been observed in experiments. Some are neutral, such as dihydride pairs and missing dimers \cite{croshaw2020atomic}. Neutral defects may still have an effect on the behavior of charges in SiDB layouts due to lattice distortion effects \cite{rashidi2018initiating, croshaw2021ionic}, but such effects are usually pronounced only when SiDBs are adjacent to those defects. Introducing lattice effects into current SiQAD ground state simulators could incur a significant performance penalty. For the purpose of logic design, the undesirable effects of neutral defects may be averted by simply avoiding the placement of SiDBs adjacent to them. A special type of neutral defect is step edges, an interface between patches on the \hsi surface where the top layer of silicon dimers is absent on one side \cite{martin1987atomic}. Their main implications on logic design are:
\begin{enumerate*}
    \item the lattice dimer orientation is rotated by $90^\circ$ on one side, meaning that logic circuits passing through step edges may require adaptors, and
    \item changes in physical parameters between objects placed across the step edge.
\end{enumerate*}
Step edges are challenging to capture in simulation and are currently not supported in any CAD or simulation tool that specializes in large-scale SiDB logic research. Therefore, they are also not considered in the current work.
On the other hand, charged defects such as near-surface dopants and silicon vacancies can exert screened Coulombic effects on SiDB logic gates at a distance \cite{huff2019electrostatic}.
Experimental fittings for three defect types were performed in \cite{huff2019electrostatic}, including:
\begin{description}
    \item[T1 arsenic] a positively-charged arsenic dopant at \SI{8}{\angstrom} depth;
    \item[T2 vacancy] an initially unclassified negatively-charged defect at \SI{4}{\angstrom} depth, later identified to be a silicon vacancy in \cite{croshaw2020atomic}; and
    \item[SiDB] a stray SiDB left over from incomplete passivation in the preparation of the surface.
\end{description}

To measure the electrostatic influence of the defects, Huff \etal{}'s employed what they denote to be a probe-SiDB as a medium to characterize the Coulombic effects from defects of interest \cite{huff2019electrostatic}. On a patch of \hsi surface containing a defect that is otherwise clean, the following procedure is performed:
\begin{enumerate}
  \item an SiDB is created near the defect to be used as a probe medium, denoted as the probe-SiDB;
  \item the tip of an atomic force microscope (AFM) is positioned above the probe-SiDB;
  \item a range of sweeping bias voltages is applied on the AFM tip and the shift in oscillating frequency is recorded, revealing a distinct charge transition of the SiDB;
  \item the probe-SiDB is erased.
\end{enumerate}
The process is repeated multiple times with the probe-SiDB placed at different distances, as illustrated in \figref{fig:exp-setup}.
The bias sweeps at the discrete distances reveal different offsets in the SiDB's charge transition levels, the Coulombic effects of the defect on the SiDB can be fitted via those offsets.

\subsection{Resilience of Logic Computation Against Variations}

 The introduction of placement and routing methods for SiDB logic has enabled the creation of SiDB circuits at unprecedented scales for this platform technology \cite{walter2022hexagons}.
In practice, such a sizeable circuit would likely encounter environmental variations which places a burden on the logic components to be resilient to these imperfections. 
Past works have studied the ability of a logic gate to abide to the intended logical behavior when subject to variations in physical parameters \cite{ng2020siqad,ng2020thes,vieira2022threeinput}. A contiguous domain of parameters within which a logic gate satisfies all truth table rows is dubbed an \emph{operational domain}.
In the screened Coulomb potential model, which is applied to experimental SiDB systems \cite{huff2018binary,huff2019electrostatic}, parameters that are relevant to the operational domain include the relative permittivity denoted $\epsilon_r$, the Thomas-Fermi screening length denoted $\lambda_\text{TF}$, and the energetic difference between the neutral/negative charge transition level and the bulk Fermi level for a lone SiDB denoted $\mu_-$. When an operational domain is plotted on the $\lambda_\text{TF}$ vs.~$\epsilon_r$ plane, changes in $\mu_-$ translate the domain along the $\epsilon_r$ axis in logarithmic scale.
However, prior studies have yet to investigate the impact of charged defects on the logical stability of SiDB gates. Whereas the operational domain captures broad changes in physical parameters that can be perceived to be uniform at the spatial scale of a logic gate, a near- or on-surface charged defect can have much more localized effects that influence each SiDB with different strengths.

\section{Methodology}\label{sec:methodology}

This section constitutes the main contribution of this work.
Taking the system energy formulation for classical ground state charge configurations as the starting point \cite{huff2018binary,ng2020siqad}, we incorporate the effects of fixed-charge defects as follows:
\begin{equation}
  E(\vec{n}) = \sum_i V_\text{ext}^{(i)} n^{(i)} + \sum_i V_\text{fc}^{(i)} n^{(i)} + \sum_{i<j} V^{(i,j)} n^{(i)} n^{(j)}
\end{equation}
where $V_\text{ext}^{(i)}$ is the sum of influences by electrostatic potential sources at the $i$th SiDB excepting those from defects and other SiDBs, $V_\text{fc}^{(i)}$ is the sum of the effects from all fixed-charge defects on each SiDB, $V^{(i,j)}$ is the screened Coulombic interaction strength between SiDBs, and $n^{(i)}$ is the charge state at each SiDB with allowed values $+1$, $0$, and $-1$ corresponding to the vacant, singly-charged, and doubly-charged states. This work adds the $V_\text{fc}^{(i)}$ term to encapsulate the Coulombic effect of defects with fixed-charges on SiDBs given by
\begin{equation}\label{eq:screened-coulomb-fc}
V_\text{fc}^{(i)} = \sum_k \frac{q_0}{4\pi\epsilon_0\epsilon_r^{(k)}} \frac{m^{(k)}}{r^{(i,k)}} \exp\left(\frac{{r^{(i,k)}}}{\lambda_\text{TF}^{(k)}}\right)
\end{equation}
where $q_0$ is the elementary charge, $m^{(k)}$ is the charge state at the $k$th defect, $\epsilon_0$ is the vacuum permittivity, $r^{(i,k)}$ is the distance between each defect and the $i$th SiDB, $\epsilon_r^{(k)}$ is the dielectric constant associated with each defect, and $\lambda_\text{TF}^{(k)}$ is the Thomas-Fermi screening length associated with each defect. To simulate these defects, the following information must be passed to a simulation engine for each defect:
\begin{enumerate*}
  \item the physical location,
  \item $z^{(k)}$,
  \item $\epsilon_r^{(k)}$, and
  \item $\lambda_\text{TF}^{(k)}$.
\end{enumerate*}
Physical parameters $\epsilon_r^{(k)}$ and $\lambda_\text{TF}^{(k)}$ are expected to be fitted experimentally as done in \cite{huff2019electrostatic} or acquired from physical simulations.
This formulation does not require alterations to established metastability conditions for SiDB charge states \cite{ng2020siqad}.

In the interest of simulation runtime, this formulation assumes that the defect being simulated maintains a fixed charge state for the physical environment in which it resides. Under extreme biases, the charged defects may take on different charge states according to the electrostatic landscape and may be subject to different metastability conditions. Capturing such effects may be of interest in future work, but is not within the scope of this formulation.

The localized band bending effects felt at the location of each SiDB can be captured by
\begin{equation}\label{eq:v-local}
    V_\text{local}^{(i)} = - V_\text{ext}^{(i)} - V_\text{fc}^{(i)} - \sum_{j} V^{(i,j)} n^{(j)}.
\end{equation}
In the presence of band bending effects, the SiDB takes a charge state that corresponds to the energetic position of its discrete charge transition levels relative to the bulk Fermi energy \cite{pitters2011charge, taucer2014singleelectron}.
When $V_\text{local}^{(i)}$ is $0$, no band bending effects are present indicating an isolated SiDB devoid of external effects.

\section{Model Verification}\label{sec:verification}

\begin{table}
  \centering
  \caption{Fitted Defect Parameters from \cite{huff2019electrostatic}}\label{tab:defect-params}
  \begin{tabular}{cccrc}
    \toprule
    Defect Type & Charge 
                       & Depth (\AA) 
                             & \multicolumn{1}{c}{$\epsilon_r$}
                                              & $\lambda_\text{TF}$ (\si{m}) \\\midrule
    T1 Arsenic  & $+1$ & $8$ & $9.7 \pm 2.5$  & $2.1 \pm 0.7$ \\
    T2 Vacancy  & $-1$ & $4$ & $10.6 \pm 0.5$ & $5.9 \pm 0.6$ \\
    SiDB        & $-1$ & $0$ & $4.1 \pm 0.2$  & $1.8 \pm 0.1$ \\\bottomrule
  \end{tabular}
\end{table}

\begin{figure}
  \centering
  \includegraphics[width=.8\linewidth]{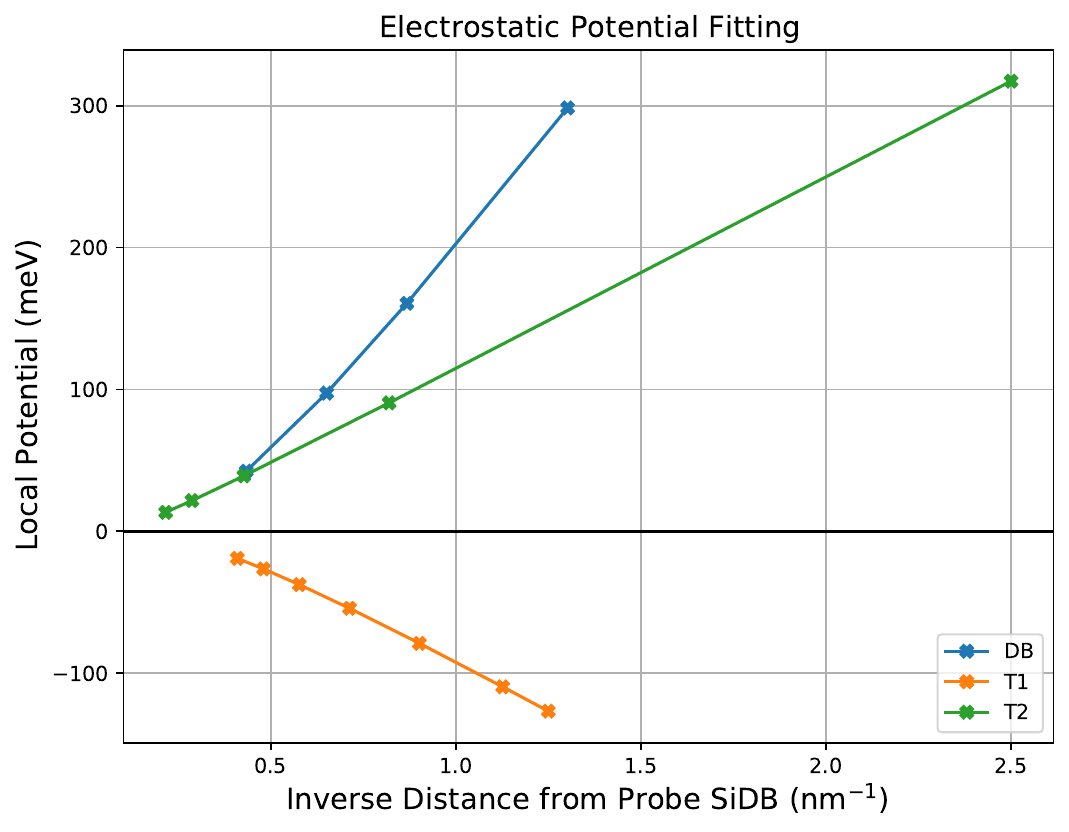}
  \caption{Simulated local band bending effects (see \eref{eq:v-local}) of the probe-SiDB at various distances when probing the three types of defects in question. The magnitude of the electrostatic effects experienced by the probe-SiDB is expected to increase as it gets closer to the defect, which is reflected in the plot as local potentials. Defect simulation parameters $\epsilon_r^{(k)}$, $\lambda_\text{TF}^{(k)}$, and $m^{(k)}$ from \eref{eq:screened-coulomb-fc} are set to those in \tabref{tab:defect-params}.}
  \label{fig:defect-coulomb-fit}
\end{figure}

To verify the correctness of the implemented formulation, we implement the model in SiQAD's SimAnneal simulator \cite{ng2020siqad} and recreate Huff \etal{}'s experiments from \cite{huff2019electrostatic} in simulation for the T1 arsenic defect, the T2 silicon vacancy defect, and an SiDB defect with parameters listed in \tabref{tab:defect-params}.
We can then compare the local band bending effects (see \eref{eq:v-local}) of the probe-SiDB between simulation and experimental fittings.
In simulation, the probe-SiDB is always configured as an SiDB object. In other words, in the simulator's point of view a probe-SiDB is no different from any other SiDB. The T1 arsenic and T2 vacancy defects are configured as fixed-charges with the corresponding physical parameters.
Note that although the expected charge state of the SiDB defect is $-1$ from the experimentally tested distances, at closer distances it may be possible to observe the SiDB defect in neutral (similar to SiDB-pairs in \cite{huff2018binary}) or positive (similar to SiDB chains in \cite{croshaw2021ionic}) states. In the simulation results presented in \figref{fig:defect-coulomb-fit}, SimAnneal returns the same results whether the SiDB defect is entered as a fixed-charge object or an SiDB object across the tested distances.
The root mean square percentage error for the T1, T2, and SiDB defects are 3\%, 17\%, and 4\% respectively from Huff \etal{}'s experimental results.
The matching results between simulation and experimental fitting confirm the simulator's accuracy in reproducing the effects from charged defects. 

\section{Application in Logic Gate Robustness Studies}\label{sec:application}

Taking advantage of the newly added defect simulation capabilities, we introduce a workflow to test a logic gate's robustness against charged defects and compare those results against their operational domains.

\subsection{Minimum Defect Clearance for Correct Logic Computation}

\begin{figure}
    \centering
    \includegraphics[width=.85\linewidth]{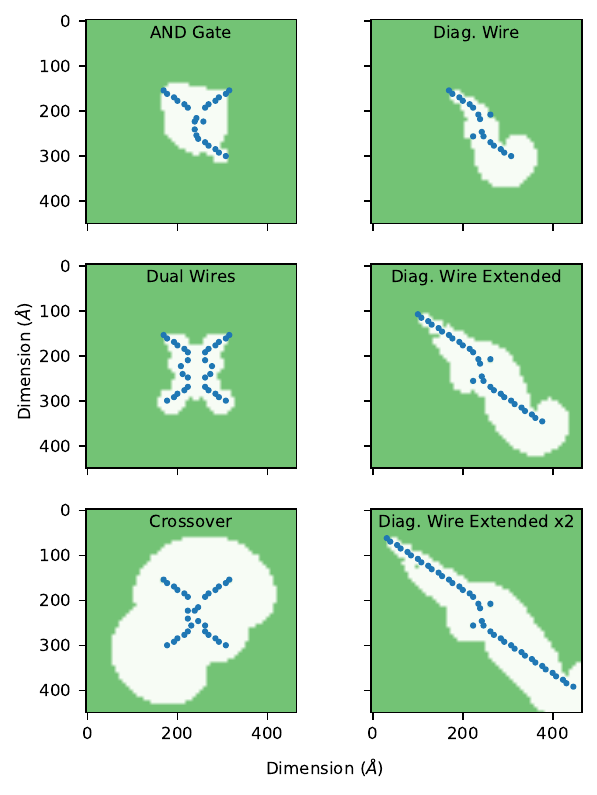}
    \caption{Defect tolerance map for select gates from the Bestagon gate library \cite{walter2022hexagons} showing locations where a single T2 vacancy defect can coexist with the gates while maintaining correct logic outputs. Green filling denotes acceptable defect locations and blue dots denote SiDB locations. The top-most SiDB-pairs in all gates are input perturbers for setting the input states per \cite{walter2022hexagons}'s design---include just the top SiDB in the logic pair for logic $0$, and just the bottom SiDB for logic $1$. The panes on the right and bottom right are modified versions of the diagonal wire tile with the input and output wires extended by different lengths. The defects are simulated within a bound of $461 \times 445 \si{\nm^2}$ centered on the gate with $\SI{7.68}{\angstrom}$ spacing between defect placements in either direction. Inter-SiDB parameters are $\mu_-=\SI{-0.32}{\eV}$, $\epsilon_r=5.6$, and $\lambda_\text{TF}=\SI{5}{\nm}$ in SimAnneal; defect parameters are taken from \tabref{tab:defect-params}. Heuristic parameters in SimAnneal are: $\text{anneal\_cycles} = 10000$; $\text{num\_instances} = 256$ as the baseline, $1024$ to verify failure thresholds, repeated as needed to reach consensus.}
    \label{fig:sim-min-exclusion-zone}
\end{figure}

When we introduce a single charged defect in the vicinity of an SiDB gate, it either continues to satisfy all truth table rows or it ceases to do so. By recording the logic correctness outcome with the defect placed at various locations, we can create a map that shows where a defect can coexist with a gate and where it causes the gate to fail. We evaluate logic correctness by inspecting the logic state of the last SiDB-pair in the output wire. In \figref{fig:sim-min-exclusion-zone}, such a map is produced showing the locations at which a T2 vacancy defect can coexist with select components from the Bestagon gate library \cite{walter2022hexagons}: AND gate, dual wires, crossover, and diagonal wire. An interesting observation between the dual wires and the crossover is that the symmetrical design of the dual wires also led to a symmetrical defect tolerance map, whereas the assymetrical crossover shows a larger clearance requirement around the left output.

For each defect location that causes a logic error, we can compute the distance between that location and its nearest SiDB to get the minimum clearance required to avoid that particular defect. Repeating this calculation for all defect locations and taking the maximum of all resulting clearance values yields the minimum defect clearance for the logic gate to remain operational for that defect type.
In \tabref{tab:min-avoid-dis-and-op-dom}, we report the minimum defect clearance for logic gates from the Bestagon gate library \cite{walter2022hexagons} in the presence of T1 arsenic at a depth of \SI{8}{\angstrom} and T2 vacancy at a depth of \SI{4}{\angstrom}. All Bestagon components are included excepting the half adder as it fails to function throughout the entire simulation grid. Across all combinations of gates and defects, the defect placement bounds and intervals as well as heuristic SimAnneal settings are consistent with those reported in the caption of \figref{fig:sim-min-exclusion-zone}. For inter-SiDB parameters, physical parameters are set to those specified by the Bestagon library \cite{walter2022hexagons}: $\mu_- = \SI{-0.32}{\eV}$, $\epsilon_r=5.6$, and $\lambda_\text{TF}=\SI{5}{\nm}$; the fixed-charge defects take the parameters from \tabref{tab:defect-params}. The planar distance is reported, meaning the distance between an SiDB and the defect's location projected onto the same plane as the SiDB. The true distance is the hypotenuse computed from the reported planar distance and defect depth. The Bestagon gate library is chosen for this study because all gates in the library share standard locations for input and output wires, which provides a fair platform for robustness comparison.
We observe that the required defect clearances are generally $<\SI{10}{\nm}$ for the Bestagon gates with a few outliers.

We further investigated the defect tolerance maps for modified versions of the diagonal wire with the input and output wires extended by 3 and 6 SiDB-pairs as shown in the right and bottom right panes of \figref{fig:sim-min-exclusion-zone}, respectively. The exclusion region tracks closely to the input wire in all variants of the diagonal wire, but a larger exclusion zone exists near the output wire. We suspect the cause to be the statically placed output perturber as opposed to the input perturbers which shift locations based on input state \cite{walter2022hexagons}. In the presence of the output perturber, a nearby negatively charged defect may only require a small bias to make logic $0$ a more likely outcome in the design of this component.
Starting from a clearance requirement of \SI{7.2}{\nm} for the original diagonal wire, at 3 and 6 SiDB-pair extensions the defect clearance increases to \SI{7.7}{\nm} and \SI{7.9}{\nm}, respectively. Note that although the exclusion region clips the border of the simulated grid in the latter case in \figref{fig:sim-min-exclusion-zone}, the same clearance value was achieved after expanding the simulation grid. We attribute the low degradation in clearance requirements to the effects of screening which exponentially attenuates Coulomb interactions. This bodes well for future investigations into multi-gate logic stability as the Coulombic effects between gates can be attenuated by placing sufficiently long wires between components.

Another interesting observation is the existence of a ``dimple'' in the white region above the bottommost SiDB-pair across all three variants of the diagonal wire, meaning that the wire is more robust against the defect in that region. We believe that the statically placed output perturber to also be the cause here, creating a sweet spot in the defect tolerance map where the negatively charged defect can exist closer to the wire than usual and still maintain correct logic outputs. If the output perturber was replaced by a longer wire or a different logic component, we expect the dimple to disappear. This result may serve as motivation for future work to consider alternative output perturber designs to more accurately capture the Coulombic effects that an output wire would exert on the logic component.


\begin{table}
    \newcommand{\colwidth}{2.8cm}
    \newcommand{\colwidthhalf}{1.5cm}
    \centering
    \caption{Defect Clearance and Operational Domain of Bestagon Gates}
    \begin{tabular}{crrrr}
        \toprule
        \multirow{2}{\colwidthhalf}{\centering Bestagon Gates \cite{walter2022hexagons}}
                    & \multicolumn{2}{c}{Defect Clearance (\si{\nm})}
                                & \multicolumn{2}{c}{Operational Domain} \\
        \cmidrule(r){2-3}\cmidrule{4-5}
                    & T1 Arsenic 
                                & T2 Vacancy 
                                            & $\mu_-$ const.      & $\mu_- \pm 10\%$  \\
        \midrule
        Wire Diag   & $ 4.1$    & $ 7.2$    & $67.0\%$      & $36.3\%$ \\
        Wire        & $ 4.5$    & $ 7.9$    & $64.5\%$      & $36.3\%$ \\
        INV         & $ 6.9$    & $12.0$    & $33.8\%$      & $11.5\%$ \\
        OR          & $ 3.3$    & $ 7.2$    & $34.0\%$      & $ 9.8\%$ \\
        AND         & $ 1.7$    & $ 6.2$    & $16.5\%$      & $ 0.0\%$ \\
        Dual Wires  & $ 3.4$    & $ 3.2$    & $14.5\%$      & $ 0.0\%$ \\
        INV Diag    & $ 5.8$    & $ 8.3$    & $13.5\%$      & $ 8.8\%$ \\
        NOR         & $ 4.9$    & $ 7.1$    & $11.5\%$      & $ 0.0\%$ \\
        Fanout-2    & $ 6.2$    & $13.7$    & $8.5\%$       & $ 0.0\%$ \\
        NAND        & $ 5.0$    & $ 6.1$    & $6.0\%$       & $ 0.0\%$ \\
        XNOR        & $ 5.0$    & $ 5.8$    & $6.0\%$       & $ 0.0\%$ \\
        XOR         & $ 5.4$    & $ 9.6$    & $4.0\%$       & $ 0.0\%$ \\
        Crossover   & $ 9.0$    & $12.9$    & $0.5\%$       & $ 0.0\%$ \\
        \bottomrule
    \end{tabular}
    \label{tab:min-avoid-dis-and-op-dom}
\end{table}

\subsection{Comparison Against Other Robustness Metrics}

\begin{figure}
    \centering
    \includegraphics[width=0.7\linewidth]{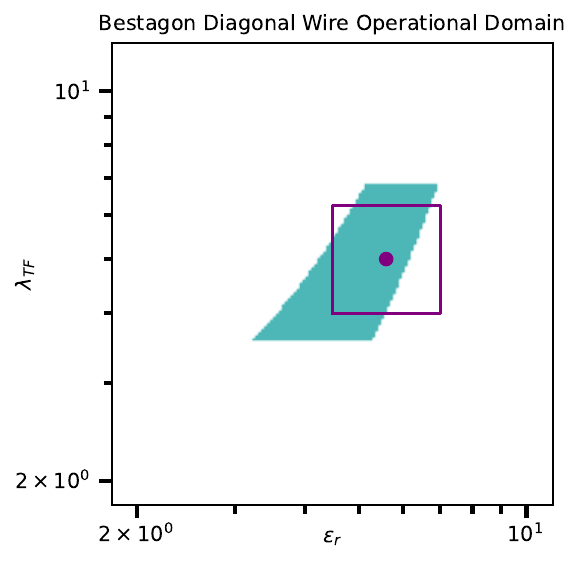}
    \caption{Operational domain of the diagonal wire tile from the Bestagon gate library \cite{walter2022hexagons} with $\mu_- = \SI{-0.32}{\eV}$ simulated with $150$ points in each dimension in logarithmic spacing. The teal-filled domain represents parameters that yield successful logic outputs for the wire; the purple point represents the $\epsilon_r$ and $\lambda_\text{TF}$ parameters that the wire was designed for; the square centered on that point represents the parameter neighborhood considered by this work which is $\pm0.25$ of the target parameters in log space. The SimAnneal heuristic parameters are set to $\text{anneal\_cycle} = 10000$ and $\text{num\_instances} = 512$.}
    \label{fig:op-dom-plot}
\end{figure}


As discussed in \secref{sec:background}, existing studies on SiDB logic have not put a major emphasis on logic stability in the presence of environmental variations, with relevant efforts mainly focused on operational domains which captures variations that affect an entire logic gate uniformly \cite{ng2020siqad,vieira2022threeinput}.
Since the effects of near-surface charged defects are much more localized, it is important to evaluate the newly proposed defect clearance against existing robustness metrics.
Past works that have studied operational domains have presented the results graphically \cite{ng2020siqad,ng2020thes,vieira2022threeinput} without offering a numeric figure of merit, necessitating an original quantification scheme for this comparison. In SiDB logic gate literature, each logic gate targets a specific set of physical parameters including $\mu_-$, $\epsilon_r$, and $\lambda_\text{TF}$. A physical \hsi specimen that is tuned to the same set of physical parameters would ideally only exhibit minor deviation across the surface. Therefore, we posit that the most relevant section of the operational domain is the nearest neighborhood around the gate's targeted parameters.
For this study, we have opted to focus on the parameter neighborhood of $\epsilon_r$, $\lambda_\text{TF}$, and $\mu_-$ that the Bestagon components are designed for. The bounds are set to $\pm0.25$ for $\epsilon_r$ and  $\lambda_\text{TF}$ in log space with a discrete parameter grid of $20$ by $20$ in log--log scale.
\figref{fig:op-dom-plot} illustrates the operational domain of the Bestagon diagonal wire \cite{walter2022hexagons} at $\mu_- = -\SI{0.32}{\eV}$, the considered parameter neighborhood is also highlighted.
The bounds for $\mu_-$ are set to $\pm0.1$ in log space as we found $\pm0.25$ to yield no working results. We believe that the chosen $\mu_-$ bounds remain to be physically relevant as the bulk dopant concentration across the sample can be prepared to be relatively stable \cite{broise2000theory}. As previously mentioned, broad variations in $\mu_-$ shifts the operational domain in the $\epsilon_r$ direction in log scale. Therefore, the operational domain with $\mu_-$ consideration can be constructed by taking the intersection of the $\lambda_\text{TF}$ vs.~$\epsilon_r$ domain at the $\mu_-$ bounds.

We quantify the robustness of each gate by taking the logical success rate within the inspected parameter neighborhood. The results are included in \tabref{tab:min-avoid-dis-and-op-dom} with the ``$\mu_-$ const.'' and ``$\mu_- \pm 10\%$'' columns showing success rates without and with $\mu_-$ variation, respectively. We observe that many gates from the Bestagon library are highly sensitive to $\mu_-$ variations, with more than half of them achieving $0\%$ success rate in the chosen $\mu_-$ bounds. This is a result of the operational domains having insufficient width in the $\epsilon_r$ axis to withstand domain translations caused by $\mu_-$ variations. This may be indicative that future logic design works should prioritize $\epsilon_r$ robustness as having higher $\epsilon_r$ tolerance also benefits $\mu_-$ tolerance.

We do not find a clear correlation between the overall trends for defect clearance and operational domain. As previously mentioned, operational domain studies apply deviations from the targeted parameters uniformly on the entire logic gate, whereas the introduction of defects exert highly localized effects thanks to the exponential effects of screening (see \eref{eq:screened-coulomb-fc}). If we inspect the local potential influences of the T2 vacancy defect plotted in \figref{fig:defect-coulomb-fit}, as the probe-SiDB moves from \SI{0.4}{\nm} to \SI{4.6}{\nm}, the Coulombic repulsion experienced by the probe-SiDB is also reduced by a factor of $>20$, showing how rapidly the defect's influence falls off.

\section{Conclusion}\label{sec:conclusion}

This work has proposed a formulation to simulate fixed-charge defects in SiDB systems, and demonstrated its accuracy by implementing it in SiQAD's ground state simulator and comparing the results with past experimental findings. The ability to simulate defects in line with experimental findings for the first time bring a valuable tool for experimentalists and logic designers alike.
Experimentalists have gained the ability to more accurately predict the behavior of SiDB layouts in the presence of charged defects and adjust the layouts accordingly before carrying out experiments. Designers may now explore the sensitivity of their circuit designs in the presence of nearby defects. Automated circuit designers \cite{lupoiu2022automated} may also make use of defect simulation capabilities to optimize gates with nearby defects in mind.
Future works have the opportunity to further add to the simulation capabilities by considering the incorporation of a dynamic charge state model that is aware of the defects' energetic charge transition levels. Metastability conditions of these dynamic defect charge states may take inspiration from the equivalent conditions proposed for SiDBs in \cite{ng2020siqad}.
The simulation of neutral defects may also be of interest, but they may involve more drastic changes to simulator implementation since the energetic effects from lattice distortions may have a dependence on the charge configuration \cite{rashidi2018initiating}, which might necessitate a re-computation for every charge reconfiguration.

In terms of evaluating the robustness of logic gate designs, this work has immediately demonstrated the methodology's application in characterizing the resilience of logic gates from the Bestagon gate library \cite{walter2022hexagons} against the introduction of a single charged defect in its vicinity, leading to a defect clearance value that can be assigned to each logic gate. 
We have also investigated the defect clearances of a diagonal wire component with extended input and output wires, observing that the defect clearance requirement did not increase significantly.
Gate library designers can use the presented methodologies to identify opportunities for further improvements.
This also creates research opportunities for the design automation community at multiple steps of the design flow. At the logic synthesis step, gates with high clearance requirements can be associated with high costs in order to reduce the use of those gates.
In the placement and routing step, awareness of defect clearance requirements can ensure that the resulting circuits are compatible with the defect landscape.

\balance

We have also studied the operational domain of the Bestagon gates by varing $\lambda_\text{TF}$, $\epsilon_r$, and $\mu_-$. We find a lack of correlation between the operational domain and the minimum defect clearance, indicating that future works that seek to study and quantify SiDB logic robustness may have to consider the interplay between broad effects captured by operational domain and local effects captured by defect clearances. We also found Bestagon gates to be very sensitive to variations in $\mu_-$, a finding that may inform future design choices.

\bibliographystyle{IEEEtran}
\bibliography{refs}

\end{document}